# Random Motion with Interfacial Contact: Driven Diffusion *vis-a-vis* Mechanical Activation

P. S. Goohpattader and M. K. Chaudhury[a]

Department of Chemical Engineering, Lehigh University, Bethlehem, PA 18015, USA

## Abstract

Rolling of a small sphere on a solid support is governed by a non-linear friction that is akin to the Coulombic dry fiction. No motion occurs when the external field is weaker than the frictional resistance. However, with the intervention of an external noise, a viscous friction like property emerges; thus the sphere rolls with an uniform drift velocity that is proportional to the applied field. As the sphere rolls, it rocks forward and backward resulting in substantial fluctuation of displacement opposite to the net drift. The ratio of the integrated probabilities of the negative to positive work fluctuations decreases monotonically with the time of observation, from which a temperature like intensive parameter can be estimated. This parameter conforms to the Einstein's ratio of diffusivity and mobility that increases almost linearly, even though the diffusivity increases super-linearly, with the strength of the noise. A new barrier crossing experiment is introduced that can be performed either with a hard (e.g. a steel ball) or with a soft (e.g. a water drop) sphere in contact with a periodically undulated substrate. The frequency of barrier crossing follows the classical transition state equation allowing a direct estimation of the effective temperature. These experiments as well as certain numerical simulations suggest that the effective temperature of a system controlled by a non-linear friction may not have a unique value.

[a] e-mail: mkc4@lehigh.edu

# 1 Introduction

This paper reports a form of a Brownian motion that is induced by a mechanical noise to a system where the friction arises from the irreversible adhesive contact of two surfaces. A random motion with an *interfacial resistance* was first discussed about fifty years ago by Caughey and Dienes [1] in the context of sliding structures responding to earthquake. Similar kinds of motion with a weak *adhesive contact* have been reported recently with a colloidal particle on a soft microtubule [2], and with a small object on a solid surface [3-5].

Frictional dynamics in many of these systems are hysteretic or non-linear [6-13], in that they are driven by instabilities [8, 9]. As Muser [9] eloquently pointed out, the viscous drag friction results from the distribution of collision energy from the central degree of freedom of a Brownian particle to other degrees of freedom of the solvent particles. However, even at a vanishingly small velocity of sliding of one solid past another, fast motions of certain degrees of freedom result in "stick slip" instability that lead to non-linear friction. These instabilities are observed not only with a spring/mass system, but with random noise excitations [5] as well. They are also observed with the relaxation of the contact line [4] of a liquid drop on a solid surface. It was proposed [14] long ago that a similar Coulomb friction like instability accompanies the collapse of the Bloch wall structures and the Barkhausen noise in magnetism as well.

Recent experiments carried out in our laboratory [3-5] showed that the sliding of a small block and the motion of a liquid drop on a solid support exhibit certain comparable characteristics in a stochastic setting. For example, when a small external force is applied, no motion occurs. However, in conjunction with an external noise, a kinematic friction like property emerges out of the static friction so that a ball moves through a granular medium [15], a slider slides [3, 5] or a drop glides [5] with an uniform drift velocity that increases linearly with the applied force. The signature of the non-linear friction, nonetheless, is evident in that the drift velocity increases non-linearly with the strength of the noise, but saturating at large values of *K* [3-5]. Furthermore, the microscopic displacement distributions are super Gaussian [3-5] at short time limit but, they all evolve towards a skewed Gaussian distribution in the long time limit. While the variance of the displacement is linear with time, the diffusivity grows super linearly with the strength of the noise. Displacement spikes [3] (stick-slip type instability) are observed as well. All these features contrast the behavior of a linear kinematic friction, where the motion is always smooth and the diffusivity grows linearly with *K*. In the current paper, we are interested to find out as to what extent such a non-linear stochastic dynamics is amenable to a standard definition of an effective temperature, e.g. the Einstein's ratio of diffusivity and mobility [16-18] or that extracted from a typical fluctuation relation [19]?

A temperature like intensive property has been long sought after [20-24] in systems driven by active as well as quenched fluctuations. In dynamic systems, ranging from vibrated granular media [19-23] to earthquake [24], various definitions of a non-equilibrium temperature have been proposed. Several path breaking experiments [21-23] were conducted as well, including a torsional pendulum immersed in a vibrated granular medium [21], fluctuation of a ball in a turbulent flow [22], and the diffusion of particles in a shear flow [23] to name a few. These experiments provided estimates of the effective temperature using the familiar concepts of statistical mechanics, such as the kinetic energy, the Einstein's ratio of diffusivity and mobility as well as the density of states [22]. Notably, Abate and Durian [22] published a paper, in which they reported reasonable agreements of the estimates of the "effective temperature" of a granular medium obtained using different metrics, mentioned as above.

Motivated by the encouraging results of the previous studies, we ask how does an "effective temperature" obtained from a driven diffusion experiment compare with an energy exchange process that we are familiar with. A sub-critical instability, such as a barrier crossing phenomenon, is an example of the latter. This subject of activated dynamics in an athermal system has also been discussed recently in the context of the deformation and flow behaviors of glassy systems [18-25], the relaxation of a sand pile [26] and the shear rate dependent stiffening of granular materials [27]. While driven diffusive experiments [3-5, 28] can be performed with various systems exhibiting non-linear friction, the systems with which to conduct both this as well as a barrier crossing experiment involve the motion of a small rigid sphere [28] on a soft fibrillated rubber substrate. The fibrillar surface mimics the features of well-decorated asperities with which a sphere undergoes a pinning-depinning [29-30] transition (fig. (2)). This leads to a threshold force somewhat like the Coulombic sliding, which has to be overcome before rolling occurs. We show below how this experiment could also be adapted to study the barrier crossing rate with the aid of an undulated support. While the bulk of our research concerns the rolling motion of a rigid sphere, we also report results of some barrier crossing experiments with a deformable sphere, i.e. a liquid drop.

## 2 Non-linear Rolling Friction

When a rigid sphere is brought into contact and separated from a fibrillated rubber surface [29, 30], a significant difference of the adhesion energy is observed signifying that the interaction of the contactor with the substrate is hysteretic. Rolling of a sphere on a surface accompanies the propagation of two cracks [31-36], one closing at the advancing edge and the other opening at the receding edge. Because of the difference in the energies of the opening and closing the cracks, a threshold force or torque is needed to roll a sphere on a substrate. Using an energy argument (see also Appendix A), one can show that a torque [31-36] of the following magnitude must be supplied about the point of contact for the incipient rolling:

$$Q \approx (W_r - W_a)\, r^2, \tag{1}$$

where, $W_r$ and $W_a$ are the receding and advancing works of adhesion or, more accurately, the strain energy release rates associated with the opening and closing the cracks and $r$ is the width of contact. The Langevin equation corresponding to the random rotation of the sphere in the presence of a stochastic force can be described by the balance of all the inertial, frictional and external torques:

$$I\frac{d\omega}{dt} + \Lambda\omega + (W_r - W_a)r^2 \frac{|\omega|}{\omega} = (\overline{F} + F(t))R, \quad (2)$$

where, $I$ is the moment of inertia of the sphere about the point of contact, $\omega$ is the angular velocity, $R$ is the radius of the sphere and $\Lambda$ is a kinematic friction factor, whereas $\overline{F}$ and $F(t)$ are the fixed and time dependent forces respectively. This equation belongs to the same class of equations proposed earlier for the sliding of a solid object [1, 37], for the motion of a liquid droplet [38] or for the friction between granular particles [39], for which several elegant solutions [40-43] are now available. The effect of a Coulombic dry friction has also been investigated [44] recently to study the kinetics of a granular asymmetric piston within the framework of a Boltzmann-Lorentz equation. In the context of sliding, it is the Coulombic dry friction while for the drop motion it is the wetting hysteresis, which are analogous to the adhesion hysteresis as outlined above. The situation of the stochastic rolling of the sphere also belongs to a class of rotational Brownian motion [45-49] that was studied earlier with a mirror hanging from a pendulum [45, 46], a floating micro-needle [47] and a colloidal sphere [48, 49]. However, in those instances, only linear friction was considered.

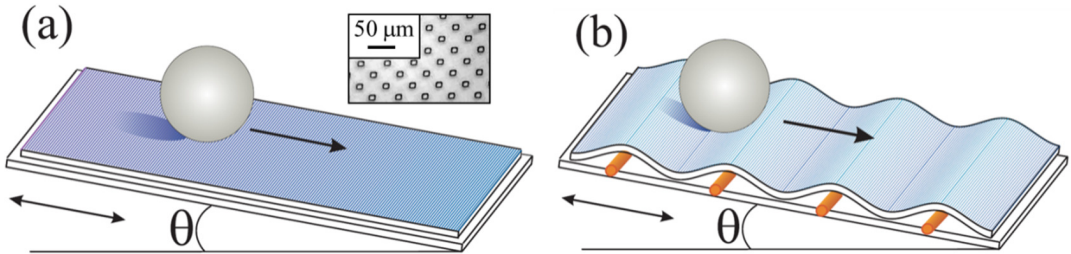

**Fig. 1**. (Colour on-line) (a) Illustration of the driven diffusive experiment with a steel ball on a fibrillated rubber surface. (b) Illustration of a barrier crossing experiment. In either case, the ball remains stationary if the angle of inclination is less than some critical angle. However, with a noise, the ball rolls down as in fig. (a) or crosses over the barrier as in fig. (b).

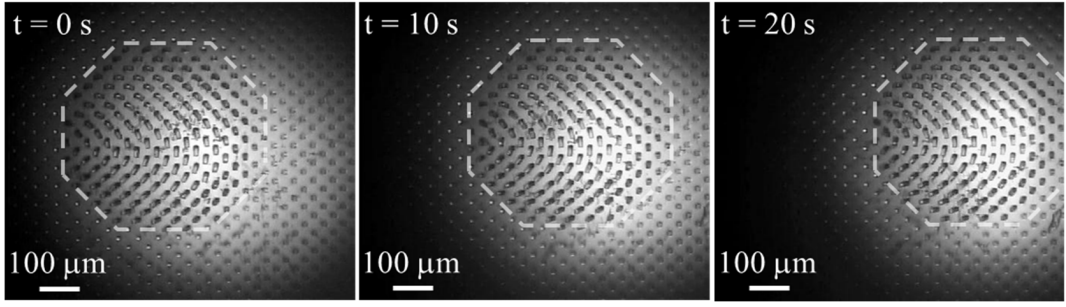

**Fig. 2.** Video microscopic images of the contact area of a steel ball rolling on a fibrillated rubber surface in the absence of noise. Here the support is slowly inclined (~ 3°) till the sphere just begins to roll. The fibrillar (dark spots) contacts are inside the dashed octagon. As the sphere rolls, the fibrils ahead of the contact make new contact with it, while those in the rear are detached. The dissipation of energy due to the relaxation of the fibrils gives rise to an adhesive hysteresis.

A translational version [28] of eq. (2) can be written down as follows:

$$\frac{7}{5}\frac{dV}{dt} + \frac{V}{\tau_L} + \sigma(V)\Delta = \bar{\gamma} + \gamma(t),$$

(3)

$$\frac{dx}{dt} = V,$$

here, $\Delta = (W_r - W_a)r^2/mR$, $\tau_L$ is the Langevin relaxation time, $\bar{\gamma} = \bar{F}/m$ and $\gamma(t) = F(t)/m$.

As has been discussed in the past [1, 28], a useful simplification of eq. (3) is to consider an equivalent linear version of this equation with a remainder term $\varphi$ shown in eq. (5). For the purpose of this discussion, we ignore the factor 7/5 of eq. (3), which is not important for the scaling argument to follow.

$$\frac{dV}{dt} + \frac{V}{\tau_L^*} = \gamma(t),$$

(4)

$$\varphi = \frac{V}{\tau_L^*} - \frac{V}{\tau_L} - \sigma(V)\Delta,$$

(5)

The criterion for equivalent linearization is to minimize the expected value of $\varphi^2$ with respect to $\tau_L^*$, which leads to the following equation:

$$\frac{1}{\tau_L^*} = \frac{1}{\tau_L} + \frac{\Delta \langle \sigma(V)V \rangle}{\langle V^2 \rangle}, \tag{6}$$

The quantities in the angular brackets of eq. (6) can be estimated with the help of a probability distribution function of the velocity, i.e. from the Fokker-Plank solution of the probability density in the velocity space. Following the procedures outlined in references [1] and [28], one obtains:

$$\frac{1}{\tau_L^*} = \frac{1}{\tau_L} + \frac{\Delta^2}{K}, \tag{7}$$

Equation (7) defines the equivalent relaxation time in terms of the Coulombic and a linear kinematic friction. When a small external bias is imposed, an expression for the drift velocity can be obtained from the linear response approximation, i.e. $V_{drift} = \bar{\gamma}\tau_L^*$. One thus has,

$$V_d = \frac{\bar{\gamma}\tau_L}{1 + \Delta^2 \tau_L / K}, \tag{8}$$

The concurrence of a randomized non-linear system to a linear response behavior as above has already been demonstrated experimentally by us in the past [3, 5]. Using the above expression for the effective relaxation time $\tau_L^*$, we can express the diffusivity as $D = K\tau_L^{*2}/2$, and an effective temperature ($T_{eff}$) as the ratio of the diffusivity and mobility as

$$T_{eff} = \frac{mK\tau_L}{2(1 + \Delta^2 \tau_L / K)}, \tag{9}$$

With a non-linear friction of the type: ($\sim |V|^n$), it can be shown [28] that the characteristic relaxation time ($\tau_L^*$) scales as $K^{\frac{1-n}{1+n}}$, so that the drift velocity, diffusivity and the effective temperature scale with the noise strength as $K^{\frac{1-n}{1+n}}$, $K^{\frac{3-n}{1+n}}$ and $K^{\frac{2}{1+n}}$ respectively. In all these cases, the effective temperature approaches a zero value more rapidly with $K$ than is the usual case with a linear kinematic friction.

## 3 Brief Summary of Previous Studies

Recently, we reported [28] the behavior of a rigid sphere on a solid support intervened by an external force and a random Gaussian noise. One main observation was that the drift velocity increases linearly at small $K$, but it saturates to a constant value at a high noise strength (fig. (3)). These results are consistent with two characteristic time scales to the problem, one being the noise independent Langevin relaxation time $\tau_L$ and the other is the noise dependent response time $K/\Delta^2$ as discussed above. In the short observation time scale, and with a weak noise, $K/\Delta^2$ dominates the drift velocity, which increases linearly with $K$. However, at large values of $K$, the dynamics is dominated by the Langevin relaxation time. This transition from a non-linear (at low $K$) to a linear control (at high $K$) of motion was further interrogated [28] by subjecting the ball to a stochastic noise and an asymmetric vibration simultaneously. At low $K$, the non-linear friction rectifies the asymmetric vibration [50], thereby giving rise to a ratchet like motion.

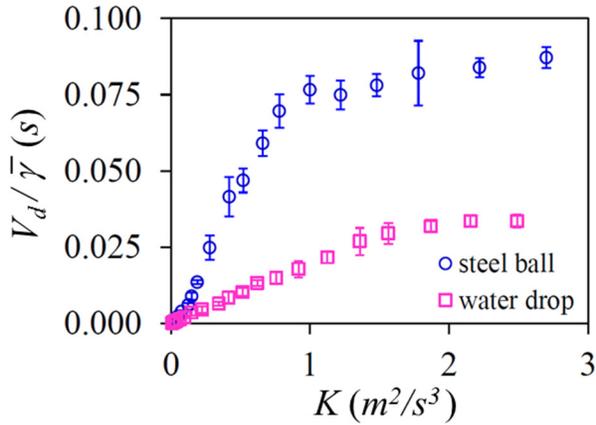

**Fig. 3**. (Colour on-line) Response times $(V_d/\bar{\gamma})$ of a steel ball (2 mm diameter) and a water drop (8 $\mu$l) rolling on the surface of a fibrillated silicone rubber as a function of the strength ($K$) of a Gaussian vibration. The data for the steel ball are from reference [28], whereas those for the water drop are from the current study.

However, with the preponderance of the linear-friction at high *K*, the drift velocity nearly vanishes. A more complex scenario of the rolling friction in the intermediate velocity range was also considered in the previous paper [28]. In particular, a super-linear velocity dependent friction plays a role in the complex evolution of the pdf of the displacement fluctuation resulting in a sigmoidal variation of the drift velocity with *K*. Specifically, a noise strength dependent state and a velocity dependent rate law was needed to explain the overall behavior of friction.

Extending the theoretical discussions of a Brownian motion to account for the sliding or rolling dynamics has certain limitations. For example, most of the theoretical frameworks of driven diffusion and barrier crossing are developed for an ideal white noise in a Markovian setting. In our experiments, all noises have finite band widths. Furthermore, the idea of extracting a temperature from the Einstein's ratio of diffusivity and mobility is sensible for systems controlled by linear friction, which are at or very close to the equilibrium. *A priori*, there is not guarantee that such a notion would apply to an athermal dynamics controlled by a non-linear friction. Our strategy here is to extract a temperature like intensive property from a driven diffusive motion of a rigid ball rolling on a surface. The novel aspect of this work is the introduction of a barrier crossing experiment from which an effective temperature can be extracted using the analogy of the theory of thermal activation. The rest of the paper is organized as follows. The first part of the paper describes the method of extracting the effective temperature from the standard method of diffusivity and mobility, or, equivalently a work fluctuation relation. Next, we venture into estimating the noise strength dependent effective temperature from the experiments of mechanical activation in the light of the Van't Hoff-Arrhenius-Eyring equation. Some discrepancy is observed in the values of the $T_{eff}$ obtained from the two methods. After discussing the possible origin of the discrepancy, we make additional conjectures.

# 4 Experimental Method

## 4.1 Drift and Diffusivity

Rolling experiments were carried out with a small steel ball (4 mm diameter, 0.26 gm, rms roughness of 35nm) on a fibrillated PDMS support that was inclined by about 1° from the horizontal and subjecting it to a random vibration (fig. 1(a)). The PDMS surface had square fibrils of 10$\mu$m size with a center to center distance of the adjacent fibrils of 50 $\mu$m. The height of the fibrils was 25$\mu$m. When a sphere rolls on a smooth rubber, the resistance to rolling [31] is amplified by the viscoelastic dissipation in the rubber. This force can be so large that very strong vibration is needed to dislodge the ball from the surface. The viscoelastic dissipation is considerably minimized on the fibrillated rubber surface owing to the diminished area of contact. The energy dissipation here is primarily due to the elastic distortion and the subsequent relaxation of the fibrils. While the adhesion/detachment processes are still hysteretic [29-30], its magnitude is low such that the ball can be easily dislodged from the surface with a small amount of vibration.

The solid support was attached to an aluminum platform connected to the stem of a mechanical oscillator (Pasco Scientific, Model SF-9324). Gaussian white noise was generated with a waveform generator (Agilent, model 33120A) and fed to the oscillator via a power amplifier (Sherwood, Model No: RX-4105). By controlling the amplification of the power amplifier, noises of different powers were generated while keeping the pulse width constant at 40 $\mu$s. The acceleration of the supporting aluminum plate was estimated with a calibrated accelerometer (PCB Peizotronics, Model No: 353B17) driven by a Signal Conditioner (PCB Peizotronics, Model No: 482) and connected to an oscilloscope (Tektronix, Model No. TDS 3012B). The pdfs of these accelerations are Gaussian with flat power spectra up to a total bandwidth of ~10kHz. The entire setup was placed on a vibration isolation table (Micro-g, TMC) to eliminate the effect of ground vibration. The motion of the ball was recorded with a high speed camera (Redlake, MotionPro, Model 2000) operating at 1000 frames/sec. Motion analysis software MIDAS 2.0 was used to track the dynamics of the steel ball. Additional details of the experiments can be found in reference [28]. The strength of noise at a given setting is nominally given as $K= \Gamma^2 \tau_c$, where $\Gamma$ is the root mean square acceleration of the stage, and $\tau_c$ is the time duration of the pulse. Even though the pdf of the acceleration pulses is Gaussian [28], there is a certain correlation of the noise pulses generated by a mechanical transducer. In a previous publication [28], we discussed this issue and showed that that the above estimate of $K$ needs to be normalized by a constant numerical factor in order for the data to be amenable to quantitative analysis. Here we do not invoke this numerical factor, as the main parameters of interest are $D$, $\mu$ and $T_{eff}$, which are all obtained directly from the driven diffusive and the barrier crossing experiments.

We perform two types of measurements. With a Gaussian noise and a bias, the object moves forward and backward randomly with a net drift [28]. At a given bias, we record the stochastic motion of the object with a high speed camera to study the trajectory over certain duration of time. The spatial segments of the trajectories corresponding to a certain observation time window are then used to obtain the distribution function (*pdf*) of the displacement fluctuation. Such a *pdf* has a given mean and a dispersion of displacements. By plotting the mean value as a function of time segment, a drift velocity is estimated. Furthermore, from the slope of the variance of the displacement versus time, we obtain the diffusivity. The ratio of this diffusivity to mobility is the first measure of the effective temperature. Estimation of the effective temperature from the displacement fluctuation is described later in the text.

### 4.2 Barrier Crossing

The barrier crossing experiment (fig.1(b)) was performed with a periodically undulated surface that was prepared by simply placing a thin (0.6 mm) rubber sheet over a flat surface decorated with parallel gold wires. By varying the diameter (25 $\mu$m to 75 $\mu$m) of the wires, barrier height was controlled. The topography of the surface produced this way is not exactly sinusoidal as the part of rubber in between the two wires makes a flat contact with the underneath surface. The overall shape is more like Gaussian humps with its height adjusted by the diameter of the wire, which are separated periodically from each other. Numerical simulation, however, shows that this difference of the undulation be it sinusoidal or periodically separated Gaussian humps, has no effect in the estimation of the effective temperature. The ball was placed in one of the valleys and then the substrate was subjected to a random noise of a given strength. The time needed for the ball to cross one of the barriers was noted with a stopwatch, or using a video camera. 35 such measurements were made at each power, from which the average escape frequency was estimated. With the substrate inclined at a given angle, average escape time was estimated at several different noise strengths with which the Van't Hoff-Arrhenius-Eyring (VHAE) plot was constructed. The number of jumps used in these experiments was optimized on the basis of simulation results so that no significant error is introduced in the averaging process.

## 5 Simulations

In order to estimate the drift velocity and the diffusivity at a given external bias and a noise strength, numerical solution of eq. (3) was carried out using a generalized integration method for stochastic differential equations [51]. Stochastic accelerations of the vibrating plate as measured using an accelerometer were used as the input, $\gamma(t)$, in the same sequence as they were generated experimentally to ensure that the noise correlation is identical in the experiment and the simulation. While the simulated drift velocity as well as the variance of the displacement did not depend on the integration time step (20$\mu$s–80$\mu$s), all the simulations were carried out with an integration step of 20$\mu$s.

Another set of simulations was carried out to estimate the barrier crossing probability. When the height of the barrier is much smaller than the spacing of the wires, as is the current case, the Langevin dynamics for the motion of the ball can be described by eq. (10).

$$\frac{7}{5}\frac{dV}{dt} + \frac{V}{\tau_L} + \sigma(V)\Delta + \frac{\pi g h}{\lambda}\cos\frac{2\pi x}{\lambda} = \bar{\gamma} + \gamma(t),$$

(10)

$$\frac{dx}{dt} = V,$$

Equation (10) was integrated by varying the strength of the noise. The trajectories thus obtained had certain numbers of discrete jumps of the ball from one potential minimum to the next. By dividing the total time of simulation with the numbers of jumps, the average escape frequency was estimated.

As discussed in the previous section, the topography of the experimental surface is not sinusoidal; it rather resembles Gaussian humps separated by regular intervals. We simulated this situation as well using a Langevin dynamics by assuming the energy potential around each hump to be $gh \exp[-(x/\sigma)^2]$ and replacing the fourth term of eq. (10) with a periodic modulation of $(2ghx/\sigma^2)\exp[-(x/\sigma)^2]$.

## 6 Results and Discussion

### 6.1 Persistence of Negative Fluctuation and Effective Temperature

At this point, we report an observation on the relationship between the effective temperature and the persistence of negative displacement fluctuations. The displacement pdfs (probability distribution function) could be fitted with a stretched Gaussian function [28], i.e. $P = P_o \exp\left\{-\left[(x-x_p)/\sigma\right]^m\right\}$ with different values of $m$ for the left and right wings of the distribution that account for the asymmetry. Here $x_p$ is the position of the peak, $\sigma$ is the width of the distribution and $P_o$ is a constant. Although these values of $m$ change with the observation time, the mean and the variance of the distribution increase almost linearly with time at each noise strength (fig. (4)). It is well-known that the determination of diffusivity from the evolution of variance suffers from poor statistics in the long time limit. Our experience with the types of system studied here and those reported in references [3] and [5] is that the variance vs time data can be fitted with a second order polynomial, the quadratic component of which is minor and, usually, decreases with improved statistics. We thus estimate the values of the diffusivity from the linear components of the variance vs time plots. The ratio of this diffusivity and the drift velocity is one estimate of the effective temperature that is compared with another estimate of it obtained from the work fluctuation, as discussed below.

The probability distribution of the displacements at a low noise strength exhibits substantial amount of negative fluctuations [28] that persists over certain observation time window. It is, therefore, tempting to analyze the data in the light of a conventional fluctuation theorem and extract an effective temperature [19] from such a plot. Unfortunately, an attempt to construct such a plot suffers from the malady that the pdfs are significantly asymmetric. Since the analysis of the displacement data in a typical fluctuation relation is carried out with the left wing of the distribution, the information contained in the right wing of any asymmetric distribution is ignored. It, perhaps, makes more sense to analyze the data using the integrated probabilities [52] of the positive and negative displacements ($P_+$ and $P_-$).

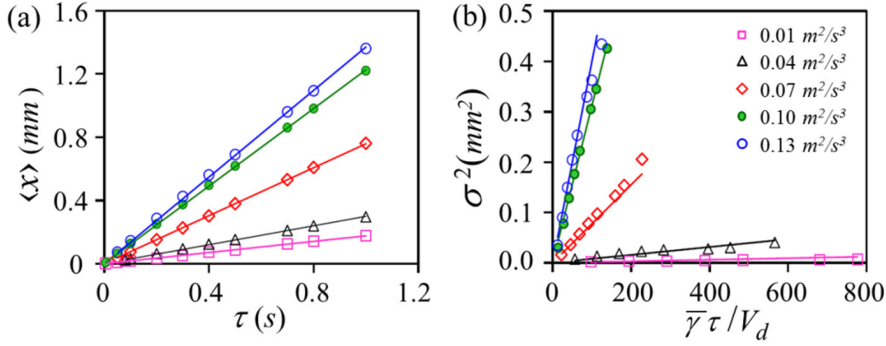

**Fig. 4**. (Colour on-line) Mean position (a) as well as the variance (b) of the displacement fluctuation of the ball increase almost linearly with time. The different symbols indicate the value of $K$ at which the data were taken.

$$P_- = P(w_\tau < 0) = \int_{-\infty}^{0} P(w_\tau) dw_t \quad \text{and} \quad P_+ = P(w_\tau > 0) = \int_{0}^{\infty} P(w_\tau) dw_t, \tag{11}$$

here, $w_\tau = m \bar{\gamma} x_\tau$ is the fluctuating work corresponding to a random displacement $x_\tau$. With the drift velocity measured at a given strength of the noise, the mean work performed on the sphere over time $\tau$ is $W_\tau = m \bar{\gamma} V_d \tau$. Figure (5(a)) shows that the ratio $P_-/P_+$ decreases monotonically with $W_\tau$. Since, there are no other variables in such a plot, the integral value of $P_-/P_+$ should be an intensive property of this driven diffusive system. When the displacement distribution is Gaussian and symmetric, the value of $P_-/P_+$ can be easily computed [53], and it can be shown that this value is only 10% higher than the ratio $D/\mu$. Figure (5(b)) shows that the temperature like intensive parameters (shown as $T_{eff}$) obtained from the integrated fluctuations and those obtained from the Einstein's ratio of diffusivity and mobility cluster very closely around the theoretical line expected for the linear friction. While such a result would be quite generic for the case with a linear friction and with a Gaussian noise, it was not anticipated *a priori* for a non-linear system (see also Appendix B). It is also gratifying to note that an estimate of the time of persistence of negative displacement fluctuation in a driven diffusive system can be obtained from this integration as: $\tau_p = T_{eff}/(m\bar{\gamma}V_d)$.

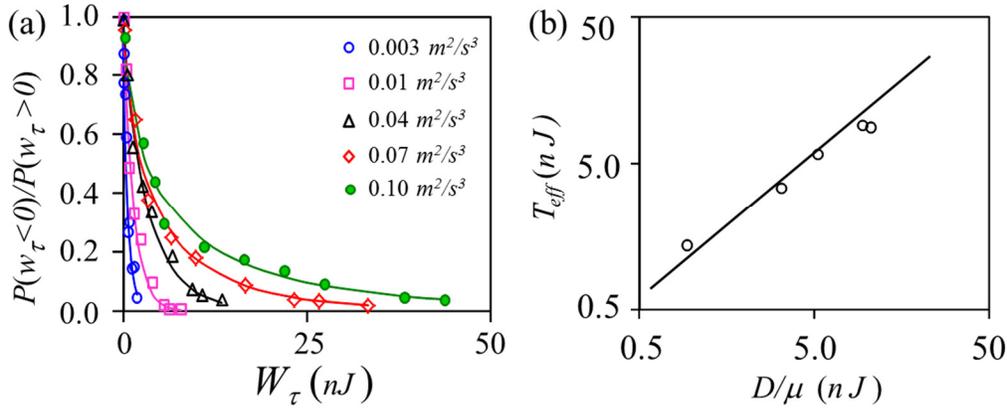

**Fig 5**. (Colour on-line) (a) An integrated work fluctuation plot for a sphere rolling on a fibrillated PDMS surface. ($P_-/P_+$) decreases monotonically with the mean work $W_\tau$ at each noise strength, $K$. All the data could be fitted with an exponential or a slightly stretched function and integrated. (b) The effective temperatures obtained from the integration of the data shown in fig. 5(a) are compared with the ratio $D/\mu$ obtained from fig.(4).

### 6.2 Effective Temperature of the Rolling Ball:

Figure (6(b)) shows that $T_{eff}$, as estimated either from the integration of $P_-/P_+$ or from the ratio of the diffusivity and mobility, increases almost linearly, even though the diffusivity increases non-linearly with $K$. This analysis yield the value of $D/\mu$ to be 90 $K$, where the former is given in terms of $nJ$ and the latter in terms of $m^2/s^3$. As the total kinetic energy of the rolling ball with a drift is about 40% higher than that of the linear kinetic energy, the effective temperature of the ball undergoing stochastic rotation should be about 126 $K\,nJ$.

At this juncture, we bring up an important issue regarding the temperature ($T$) of a non-equilibrium system as was pointed out by Speck and Seifert [54]. These authors showed that the diffusivity ($D$) of the particle in a non-equilibrium driven diffusive state is always larger than $T\mu$ by a certain amount. The subject has recently been re-iterated by Chaudhuri and Chaudhuri [55], who calculated the values of $D$ and $T\mu$ using a flashing ratchet model and reported that the ratio $D/T\mu$ departs from the equilibrium value of unity as a function of the asymmetry of the ratchet. Although it is premature to adapt these analyses to our system governed by a non-linear friction, we are led to suspect that $D/\mu$ could be an over-estimated value of the actual temperature. This parameter should now be treated as an apparent temperature, which must be compared with a value obtained from a more direct measurement (see below).

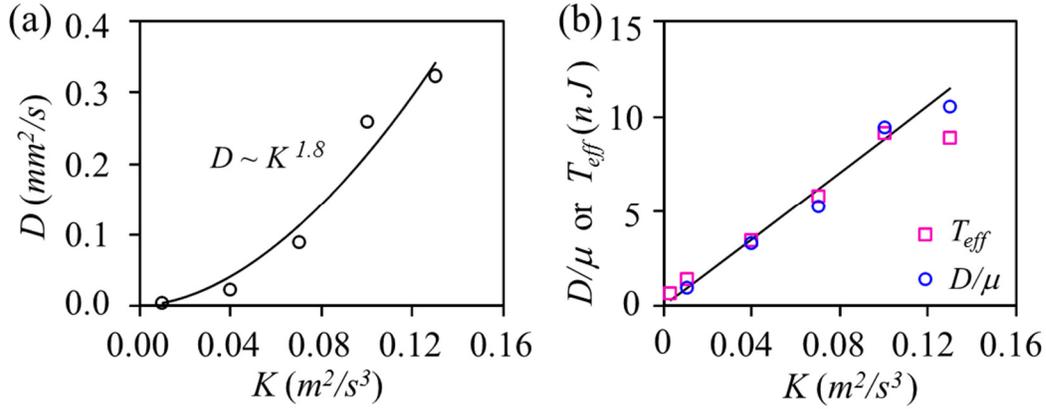

**Fig. 6.** (Colour on-line) (a) The diffusivity of the sphere increases non-linearly with the strength of the noise. Because of the limited set of data obtained within the range of $K$ studied, the fitted line is forced to go through the origin. (b) $D/\mu$ increases almost linearly with $K$. The pink squares correspond to the effective temperatures obtained from the integration of the data shown in fig. 5(a). At the lowest power, diffusivity could not be estimated accurately because of the scatter in the data, although a $T_{eff}$ could still be estimated from an integrated fluctuation plot. The data are not well-behaved at $K > 0.1$ $m^2/s^3$. All the barrier crossing experiments were carried out with $K < 0.1$ $m^2/s^3$.

## 6.3 Barrier Crossing and Van't Hoff-Arrhenius-Eyring Equation

As described in the experimental section, the barrier crossing experiment (fig. 1(b)) was performed with a steel ball on a periodically undulated rubber substrate, in which the amplitude of the undulation was varied from 25 $\mu$m to 75 $\mu$m. At relatively larger barrier heights (i.e. 50 $\mu$m and 75 $\mu$m), high noise strengths (1 $m^2/s^3$ to 4 $m^2/s^3$) were required to initiate barrier crossing as compared to the low noise strengths (0.01 $m^2/s^3$ to 0.1 $m^2/s^3$) required for a barrier height of 25 $\mu$m. We have already noted that the friction becomes linear for $K > 1$ $m^2/s^3$, where the frequency of barrier crossing increases with the noise strength as well as the tilt angle of the substrate that reduces the energy barrier.

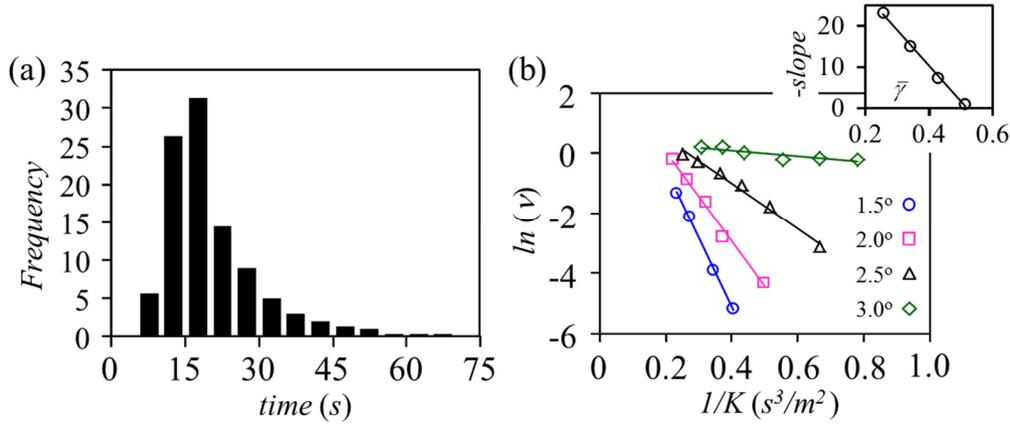

**Fig. 7**. (Colour on-line) (a) A typical distribution of waiting times of the ball before it crosses from one potential valley to the next. (b) VHAE type plots obtained with a barrier height of 75 $\mu$m at different angles of inclination. As the angle of inclination increases, the barrier height decreases leading to a diminished slope of the VHAE line. The inset shows that the slopes of these lines as a function of the bias ($\bar{\gamma}$).

The escape rate follows the rudimentary form of a force activated [56-59] Eyring's equation (Van't Hoff-Arrhenius-Eyring or VHAE form) as follows:

$$\nu = \nu_o e^{-\frac{(gh - \bar{\gamma}\lambda)}{K\tau^*}},$$

(12)

here, $\nu$ is the rate of escape, $\lambda$ is an activation length and $\tau^*$ is a time scale that converts the noise strength to an effective temperature as $mK\tau^*$. The experimentally measured escape frequencies conform well to eq. (12), i.e the plots of $\ln(\nu)$ vs $1/K$ are linear. Furthermore, the slopes of these lines vary linearly with $\bar{\gamma}$ (fig. 7(b)) from which $mK\tau^*$ is estimated to be $4K$ nJ, which is remarkably same with the experiments performed with the 50 $\mu$m and 75 $\mu$m height barriers. An estimation of the effective temperature using the method of displacement fluctuation, however, could not be conveniently performed at high $K$, where the ball exhibits rather fast dynamics. These analyses could, however, be performed comfortably at a low noise strength.

## 6.4 Barrier Crossing with Non-Linear Friction

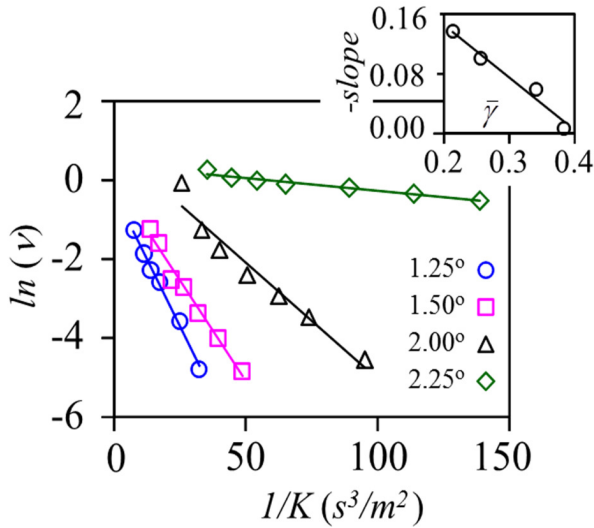

**Fig. 8**. (Colour on-line) VHAE type plots obtained with a barrier height of 25 $\mu$m at different angles of inclination. The inset shows the slopes of these lines as a function of the bias ($\bar{\gamma}$).

The results of the barrier crossing experiment at a barrier height of 25 $\mu$m are summarized in fig. (8). The logarithm of the barrier crossing rate is still linear with $1/K$, even though these low-$K$ dynamics are controlled by a non-linear friction. We already got the hint that this could be so, as the ratio $D/\mu$ varies almost linearly with $K$. From the slopes of the $\ln(\nu)$ - $1/K$ plots (fig. (8)) an effective temperature is estimated to be $T_{eff} = 208K$ nJ. This value is nearly 60% higher than that obtained from the driven diffusive experiments. In order to shed more light on this discrepancy, we simulated barrier crossing experiments by integrating eq. (10) first with a linear friction model and then with a non-linear friction model assuming certain values of $\Delta$ and $\tau_L$ [28].

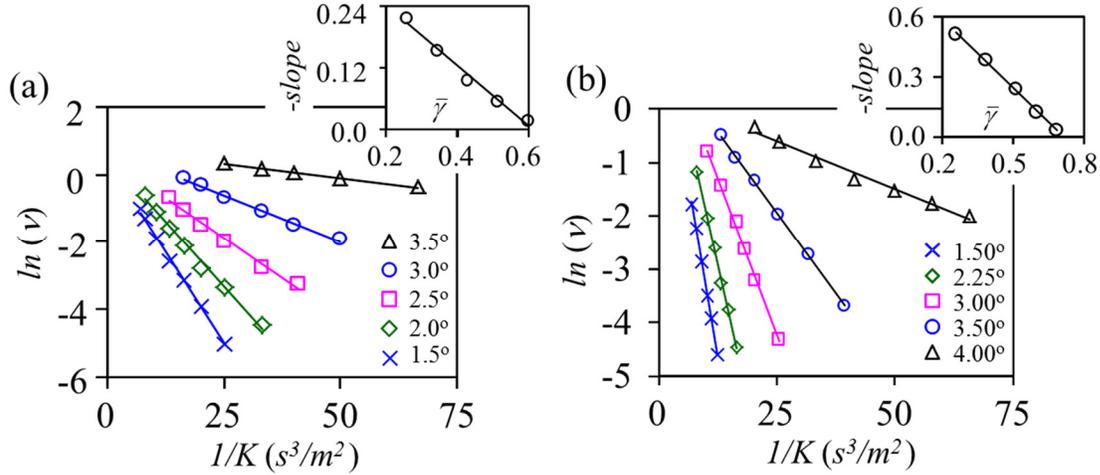

**Fig. 9**. (Colour on-line) (a) VHAE plots simulated with a linear friction model i.e. eq. (10) with $\Delta=0$ m/s$^2$ and $\tau_L$=0.01s. Barrier height of 25 $\mu$m and periodicity $\lambda$ of 1 mm is used for simulation at different angles of inclination shown inside the figure. The slopes of these lines are plotted as a function of $\bar{\gamma}$ in the inset of the fig. 9(a). (b) Similar plot as in (a) except that a non-linear friction model was used, i.e. eq. (10) with $\Delta=0.8$ m/s$^2$ and $\tau_L$=0.1s. While all the data were obtained with a surface having a sinusoidal profile, identical values of $T_{eff}$ were also obtained (not shown here) with a surface having Gaussian humps separated at same periodic intervals as $\lambda$.

The barrier crossing simulations performed with a linear friction model (i.e. $\Delta=0$) is consistent with eq. (12) in that the plots of ln($v$) vs $1/K$ are linear at all angles of inclinations. Simulations of diffusivity and mobility yielded the value of $D/\mu$ to be 100 $K$ nJ. By correcting for the rotational motion, the effective temperature is estimated to be about 140 $K$ nJ, which is slightly smaller than that (175 $K$ nJ) obtained from the barrier crossing experiment. Simulations with the combination of a linear and a non-linear friction show that the drift velocity increases linearly, but the diffusivity increases super-linearly with $K$ leading to $D/\mu \sim K^{1.5}$. Barrier crossing frequencies (fig. 9(b)) obey eq. (12) in that the plots of ln($v$) vs $1/K$ are linear at all angles of inclination. Consequently, the corresponding effective temperature is linear with $K$, i.e. $T_{eff} = 80K$ nJ, which differs from what is observed with the ratio of diffusivity and mobility (i.e. $T_{eff} \sim K^{1.5}$) (fig. (10)). We thus conclude that the two methods do not yield the same effective temperature.

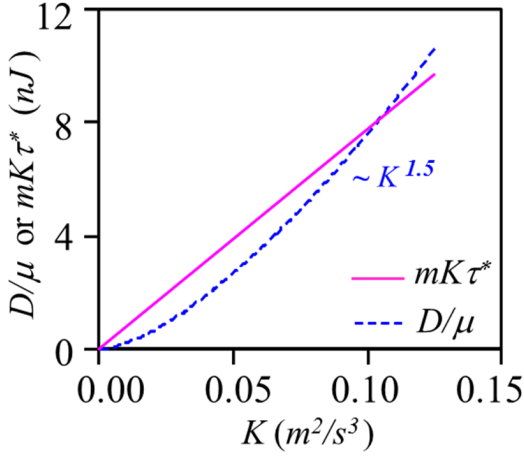

**Fig. 10**. (Colour on-line) Comparison of the $D/\mu$ and the $mK\tau^*$ values as obtained from the barrier crossing simulations with a non-linear friction model.

While the simulation reproduces the general experimental features that the diffusivity is super-linear and the drift velocity is linear with the noise strength, it has not been able to reproduce the fact that $D/\mu$ is linear with $K$. Clearly, a much more detailed state and rate dependent model of rolling friction would be needed in order to make better comparison between the simulation and the experiment. Nevertheless, both the experiment [60] and the simulation suggest that the effective temperature of a non-linear system does not have a unique value. As already pointed out, there are two response times ($\tau_L$ and $K/\Delta^2$) to a non-linear dynamics. The effective temperature may then be determined by any of those values or by their average (eq. (9)) depending upon the experiment used to interrogate it. It is plausible that the diffusivity is biased by the slower part of the dynamics, the response time of which is dominated by $K/\Delta^2$, whereas the barrier crossing is dominated by the higher end of the velocity distribution that is dominated by $\tau_L$. In other words, the ball could be hotter at the transition time scale than the overall diffusive time scale.

Another interesting part of the story is that the effective temperature (~ 20 nJ) at $K \sim 0.1$ m$^2$/s$^3$ is found to be considerably higher than that (4 nJ) at a value of $K \sim 1$ m$^2$/s$^3$. This surprising result suggests that only a small part of the externally supplied energy is transmitted to the ball at high noise strength. This would be possible if the ball spends a considerable time in a levitated state, i.e. detached from the rubber support. This picture is, in fact, supported by the video microscopic observations. No detachment of the ball, however, occurs with a low strength of the vibration. The ability of the ball to cross a larger barrier at a higher $K$ with a reasonable rate, in spite of a reduced $T_{eff}$, can be ascribed to a reduced friction, thus to the enhancement of the pre-exponential factor of the VHAE equation. This transition from an attached to a partially levitated (i.e. detached) state also appears to be a reason of the solid-like to a fluid-like transition of the drift velocity (fig. (3)) that is observed in going from a low to a high strength of the vibration.

### 6.5 Barrier Crossing Experiment with Water Drops

We conclude this section by reporting a barrier crossing experiment with a soft deformable sphere, such as a drop of water, which, in the same spirit of the above section, exhibits a slowing down of barrier crossing rate due to a difference in $v_o$. The origin of the non-linear friction here is the wetting hysteresis [5] that gives rise to a sub-linear growth of the mobility of the drop as a function of the strength of the noise (fig. (3)). The results of the experiments performed with small droplets (8 $\mu$l) of water on an undulated rubber surface (barrier height of 25 $\mu$m) are summarized in fig. (11). While the slopes of the VHAE plots with a water drop are nearly parallel to those of a rigid sphere - meaning that the effective temperatures normalized by the masses of the respective objects are the same in both cases, the pre-exponential factor ($v_o$) for the water drop is nearly half of that of the steel sphere.

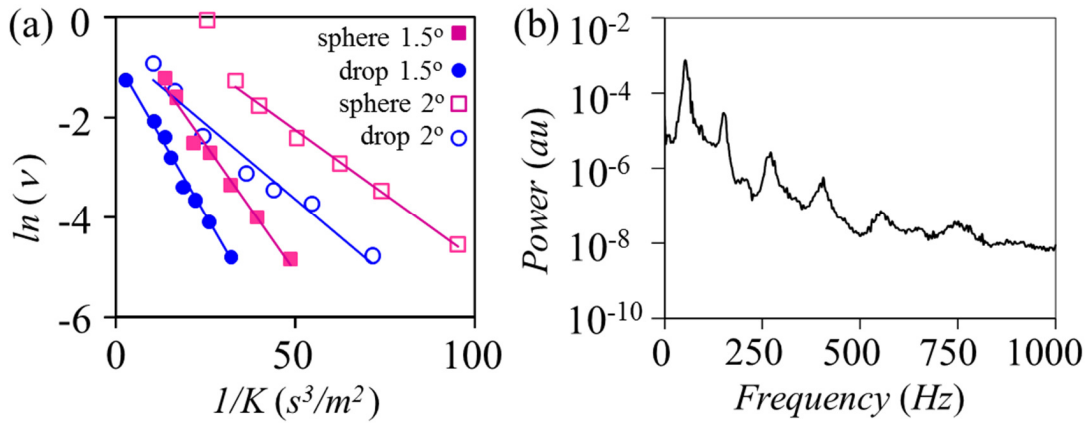

**Fig. 11.** (Colour on-line) (a) Comparison of the VHAE plots obtained for a sphere and a drop of water with a barrier height of 25 $\mu$m. The data for the sphere are same as those summarized in fig.(8). (b) A sessile drop exhibits shape fluctuation when it is excited with a Gaussian noise. Various harmonics of the shape fluctuation are shown in this power spectrum that was obtained by averaging several power spectra and de-noising it with a wavelet transform in order to reduce the background noise.

As the pre-exponential factor is inversely proportional to the frictional relaxation frequency [61] in the Smoluchowski limit, one may say that the velocity relaxation rate of a water drop is greater than that of the steel ball. This trend is consistent with the fact that the slope of the mobility versus $K$ of the water drop (fig. (3)) on a flat surface is nearly half of that of the steel sphere. By the same token, one would expect that the effective temperature of the water drop to be smaller than the rolling steel ball, which is, however, not the case. The dynamics of a liquid drop is richer than that of a steel ball in that it undergoes a noise induced oscillation composed of numerous spherical harmonics [62] (fig.11(b)). Further experiments with liquids drops of different surface tension and viscosity could shed more light on whether these noise induced oscillations could contribute to the effective temperature of the drop in the barrier crossing experiments.

## 7 Concluding Remarks

We presented here a report on the rotational coupled with a translational behavior of a small sphere on a surface that is driven by external fields and randomized by external noises. The results are encouraging in that an intensive temperature like parameter could be estimated from the decay of the negative displacement fluctuation, which is consistent with that obtained from the Einstein's ratio of diffusivity and mobility. With such a model, it was also possible to design a novel barrier crossing experiment, thus allowing measurements of the escape frequency of the sphere in terms of the tilt angle of the substrate and the barrier height. The overall behavior is consistent with the Van't Hoff-Arrhenius-Eyring form of the escape rate in its rudimentary form. The results with a small barrier height could be analyzed in detail as the dynamics was slow enough to be followed carefully. This region is also interesting owing to the fact that the dynamics is controlled by a non-linear friction that renders the diffusivity to be super-linear in *K*. While an effective temperature could be estimated from the barrier crossing experiments, a discrepancy, nonetheless, has been observed between this estimate and that obtained from *D/μ* in the low *K* region. It is plausible that these two experiments may be probing different regions of the velocity statistics of a non-linear system that exhibits two different response times. We expect that much can be learned on this issue by performing barrier crossing experiments with a fat tailed (e.g. a stretched exponential) noise that could accentuate any difference of the type mentioned as above.

## Appendix A. (Adapted from Greenwood *et al*.[31])

The origin of the non-linear friction in eq. (1) can be understood rigorously in a 2d system, with a cylinder rolling on a rubber slab in which the contact is smooth and rectangular. According to the theory of contact mechanics [31-33], the stress distribution underneath the cylinder is:

$$\sigma(x) = (b^2 - x^2)^{-1/2} \left( -\frac{P}{\pi} + \frac{E^*}{2R}(x^2 + xd - 0.5b^2) \right), \quad (A.1)$$

Here, *b* is the half-width of contact band, *R* is the radius of the roller, *P* is the applied load, *d* is the shift of the midpoint of the contact band from the point beneath the roller center, and *E\** is the contact modulus. According to the Griffith-Irwin theory of fracture mechanics, the stress has a square root singularity near the contact edges. The corresponding stress intensity factors are ($x \to \pm b$):

$$N\big|_{\pm b} = \frac{1}{\sqrt{2b}} \left( -\frac{P}{\pi} + \frac{E^*}{2R}(0.5b^2 \pm bd) \right), \quad (A.2)$$

Or, $\quad \dfrac{4RN}{E^*}\bigg|_{\pm b} = \dfrac{1}{\sqrt{2b}} \left( -b_h^2 + b(b \pm 2d) \right),$

Using the relationship between the strain energy release rate, the stress intensity factor and the contact modulus $G = \pi N^2 / E^*$, we have

$$\frac{32}{\pi} \frac{R^2 G}{E^*}\bigg|_{\pm b} = \frac{1}{b}\left(b^2 - b_h^2 \pm 2bd\right)^2, \tag{A.3}$$

Subtraction of the values of $G$ at the receding and advancing edges gives

$$[G|_{+b} - G|_{-b}] = \frac{\pi}{4} \frac{E^*}{R^2} d(b^2 - b_h^2), \tag{A.4}$$

The rolling torque $Q$ is obtained from the integration of the stress distribution as:

$$Q = dL\left(\pi E^* b^2 /(4R) - P\right) = L\left(\pi E^* /(4R)\right)(d)(b^2 - b_h^2), \tag{A.5}$$

Combining eq. (A.4) and (A.5), we have the essential result:

$$Q = RL[G|_{+b} - G|_{-b}], \tag{A.6}$$

Although eq. (A6) is derived here using the method of contact mechanics for a smooth contact, it can also be derived entirely using an energy argument [31-36]. This energy argument is also applicable for the sphere on the fibrillated surface either for a circular or for a symmetric polygonal contact. If we equate the strain energy release rate with the work of adhesion in making and breaking the contact, we can write down an equivalent expression for the rolling torque of a sphere on a flat surface as:

$$Q \approx [W_r - W_a]r^2, \tag{A.7}$$

where $r$ is the width of contact.

## Appendix B

The effective temperature obtained from the integral of $P_-/P_+$ with $W_\tau$ has also been tested against the ratio of the diffusivity and mobility for several other non-linear systems. One of those involves the sliding of a small glass cube on a slightly inclined (2°) glass plate in the presence of a Gaussian [3] or a non-Gaussian noise [5]. The second and the third cases involve the motion of a small water drop on a smooth silicon wafer induced either by a chemical [5] or a thermal gradient [63] of surface energy in the presence of a Gaussian noise.

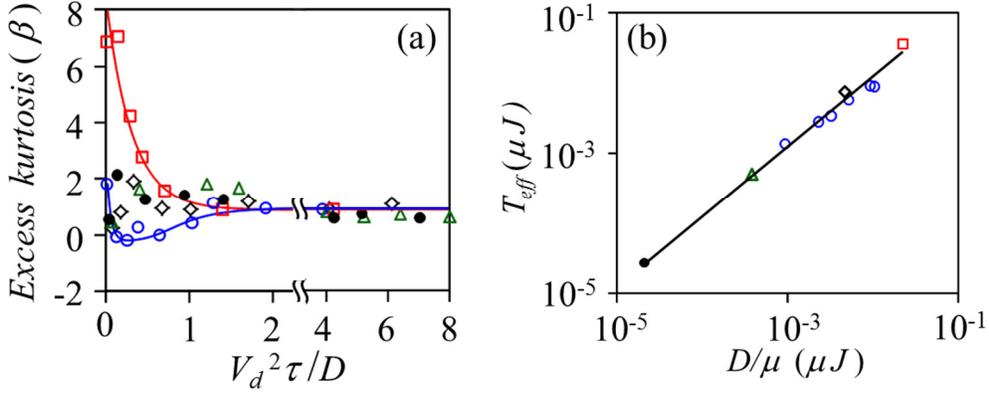

**Fig. 12.** (Colour on-line) (a) Excess kurtosis ($\beta$) is plotted against dimensionless time ($V_d^2\tau/D$) for some representative cases. (b) $T_{eff}$ as a function of $D/\mu$ for different systems. Black diamond represents the sliding (2° inclination) of a glass cube on glass surface excited by Gaussian noise, red square depicts same system excited by stretched exponential noise, blue open circle corresponds to rolling sphere on fibrillated PDMS surface subjected to Gaussian noise (all the data are from the current work, except one from a previously published work [28]), green triangle represents water drop on wettability gradient surface and filled black circle depicts water drop on thermal gradient surface.

With the measured diffusivity and the drift velocity as discussed in the text, we define a non-dimensional observation time for each system as $V_d^2\tau/D$. The first observation we make is that the kurtosis of the distribution is not constant with respect to the time of observation (fig. (12a)). Although a Leptokurtic distribution is observed at short observation time, the excess Kurtosis plateaus out to a constant value in all cases. While different evolutions are observed with different systems, the effective temperature obtained from the integrated fluctuation is in good agreement (fig. (12b)) with the Einstein's ratio of diffusivity and mobility in all these cases that include (rather surprisingly) the sliding induced by a highly stretched exponential noise [$P = P_o \exp\{-(\gamma/\sigma)^{0.3}\}$]. These results provide additional support to that discussed in the text, namely that the integration of the $P-/P+$ with respect to $W_\tau$ is a temperature like intrinsic property, i.e. it is equal to $D/\mu$.

*The help of Prof. Jagota and his group on the preparation of the fibrillated rubber is gratefully acknowledged.*